\documentclass{aa}
\usepackage[varg]{txfonts}
\usepackage{psfig,graphicx,natbib, gensymb}

\begin{document}

\title{A cautionary tale: limitations of a brightness-based spectroscopic approach to chromatic exoplanet radii}
\titlerunning{Limitations of a brightness-based approach to chromatic exoplanet radii}
\authorrunning{Cegla et al.}

\author{H.~M. Cegla\inst{1,2, 3}
	\and C. Lovis\inst{2}
	\and V. Bourrier\inst{2} 	
	\and C.~A. Watson\inst{1}
	\and A. Wyttenbach\inst{2}
} 

\offprints{H. M. Cegla, \email{h.cegla@unige.ch}}

\institute{Astrophysics Research Centre, School of Mathematics \& Physics, Queen's University Belfast, University Road, Belfast BT7 1NN, United Kingdom
  \and Observatoire de Gen\`{e}ve, Universit\'{e} de Gen\`{e}ve, 51 chemin des Maillettes, 1290 Versoix, Switzerland
  \and Swiss National Science Foundation NCCR-PlanetS CHEOPS Fellow
}

\date{Received 31 10 2016 / Accepted 23 12 2016}

\abstract{Determining wavelength-dependent exoplanet radii measurements is an excellent way to probe the composition of exoplanet atmospheres. In light of this, \cite{borsa16} sought to develop a technique to obtain such measurements by comparing ground-based transmission spectra to the expected brightness variations during an exoplanet transit. However, we demonstrate herein that this is not possible due to the transit light curve normalisation necessary to remove the effects of the Earth's atmosphere on the ground-based observations. This is because the recoverable exoplanet radius is set by the planet-to-star radius ratio within the transit light curve; we demonstrate this both analytically and with simulated planet transits, as well as through a reanalysis of the HD\,189733\,b data.
}

\keywords{Methods: data analysis -- Planets and satellites: atmospheres -- Planets and satellites: fundamental parameters -- Planets and satellites: HD\,189733\,b -- Techniques: radial velocities -- Techniques: spectroscopic} 

\maketitle

\section{Introduction}
\label{sec:intro}
Transmission spectroscopy is an essential tool for characterising the atmospheres of transiting exoplanets \citep[see e.g.][and references therein]{charbonneau02, pont13,madhusudhan14, sing16}. \cite{snellen04} demonstrated that narrowband exoplanet features (e.g. sodium) could be probed by analysing the shape of the stellar absorption lines as a planet occults its host star, that is, by studying the chromatic Rossiter-McLaughlin (RM) effect; more recently \cite{digloria15} have shown that this effect can also be used to probe broadband signatures (e.g. Rayleigh scattering). Recently, \citet[][hereafter B16]{borsa16} presented a new technique using (a version of) line-profile tomography, with the intent of studying chromatic changes in planetary radii. However, we demonstrate in this Letter that the technique in B16 is unfortunately flawed.

In principle, the application of line-profile tomography is well motivated for exoplanet atmosphere characterisation if correctly implemented. This technique isolates the starlight behind the planet during transit \citep{cameron10}, and the ratio of the integrated flux within the local profile (behind the planet) to the out-of-transit profile is equal to the brightness of the occulted starlight. In turn, the ratio of the occulted starlight is dependent on the planet-to-star radius ratio. As such, if one had space-based spectra then the planet radius could be recovered from the spectra alone, and doing so in various passbands would characterise the planetary radius wavelength dependency. For ground-based spectra this is not possible due to, for example, transparency variations of the Earth's atmosphere. To remove these effects, ground-based spectra must first be continuum normalised; they can then be multiplied by a transit light curve to allow one to study the local, occulted stellar profiles directly by subtracting the in- from out-of-transit observations (see, e.g. \citealt{cegla16b}). Since the transit light curve is dependent on the planet-to-star radius ratio, it sets the planet radius that is recoverable when examining the brightness ratio between the local and out-of-transit profiles (meaning one cannot recover new radius variations following B16). 

In this study, we first break down the physical implications of the B16 technique, and then simulate the planet transit of HD\,189733\,b to illustrate the impact of the technique's shortcomings on the measured planet radius; we also reapply the B16 technique to HARPS data, with a more rigorous error propagation. In Sect.~\ref{sec:calc}, we demonstrate how the choice of the transit light curve normalisation sets the recoverable planet radius. We present the simulated planet transits, and our reanalysis of the HARPS data in Sect.~\ref{sec:sim_recover}, and show how an underestimation of the errors can lead to spurious claims of planetary radius variations. Finally, we summarise our conclusions in Sect.~\ref{sec:conc}.

\vspace{-15pt}
\section{Limitations of a brightness-based approach to chromatic exoplanet radii}
\label{sec:calc}
In B16, the authors attempted to study passband-dependent planet radius variations by averaging together the cross-correlation functions (CCFs) for subsets of HARPS spectral orders. All of these CCFs, regardless of passband, were continuum normalised and then further scaled using a \cite{mandel02} transit light curve based on the system parameters determined in the full HARPS passband, except for passband dependent limb darkening coefficients.

For each passband, they created master out-of-transit CCFs by averaging together all the individual out-of-transit CCFs in a given night, and then subtracting the out- from the in-transit data to obtain CCFs for the starlight occulted by the planet. The authors then fitted Gaussian functions to the respective CCFs, presumably to act as a proxy for the integrated flux within, and argued that the ratio between the areas (determined from the Gaussian fits) of the local to the out-of-transit CCFs could serve as a measure of the brightness of the missing starlight, ~$\beta$, occulted during the in-transit observations.

The authors then compared this empirical $\beta$ to the expected brightness ratio based on an approximation to the solution for integrating the starlight behind the planet given the particular geometry of the system (and assuming a particular function for the limb darkening). Since this solution was dependent on the planet radius, the authors argued that they were able to disentangle a planet radius measurement for each passband. 

We demonstrate in Sect.~\ref{subsec:tran} that the technique outlined by B16 can only recover the planetary radius set by the transit light curve used in the initial normalisation. However, we emphasise that such a limitation is set due to a flux-based approach and would not be present in an RV-based approach, such as that in \cite{digloria15} or that in traditional line profile tomography (i.e. following the \citealt{cameron10} formulation). 

\vspace{-10pt}
\subsection{Transit light curve normalisation}
\label{subsec:tran}
B16 started by determining the area of their out-of-transit master CCFs, $A_{out}$, for each passband. Since the out-of-transit master CCFs ($CCF_{out}$) were continuum normalised their area is equal to their equivalent width, $EW_{out}$, that is,
\begin{equation}
\label{eqn:Ao}
A_{out} = EW_{out}.
\end{equation}  
The authors also measured the areas of the local CCFs behind the planet, $A_{loc}$. Since the in-transit CCFs ($CCF_{in}$) are normalised by the transit light curve, the area of the local CCF ($CCF_{loc})$ is
\begin{equation}
\label{eqn:Al}
A_{loc} = (1-f_{lc}) \ EW_{loc} = \beta_{lc} \ EW_{loc},
\end{equation}
where $f_{lc}$ is the flux from the light curve, $EW_{loc}$ is the equivalent width of the local CCF, and $\beta_{lc}$ is the fraction of starlight occulted by the planet under the assumptions of the transit light curve used in the normalisation. 

In an ideal case, where the local stellar photospheric profiles can be represented by constant Gaussian functions (assumed both above and in B16), then the only difference between the local and disc-integrated out-of-transit CCFs is the broadening by stellar rotation present in the disc-integrated observations. Since the rotational broadening preserves the equivalent width, then 
\begin{equation}
\label{eqn:ew}
EW_{out} = EW_{loc} = EW_{in},
\end{equation} 
where the $EW_{out}$ and the $EW_{loc}$ are not only equal to each other but are also equal to the in-transit equivalent width, $EW_{in}$ (since it is simply a summation of local profiles of equal EW), and the ratio of the areas becomes:
\begin{equation}
\label{eqn:Arat}
\frac{A_{loc}}{A_{out}} = \beta_{lc}.
\end{equation}

Hence, under the above assumptions one can only recover the planet radius injected into the model transit light curve, regardless of which passband is studied. This is because the transit light curve normalisation effectively sets the area of the local profile. We note, that the transit light curve normalisation for the above follows \citet[][hereafter C16]{cegla16b}, where $CCF_{loc} = CCF_{out} - f_{lc}*CCF_{in}$; whereas, the transit light curve normalisation in B16 was
\begin{equation}
\label{eqn:Bnorm}
CCF_{loc}^{B16} = CCF_{in} \ f_{lc} - CCF_{out} + (1-f_{lc})\end{equation}
(Borsa, Private Comm.), such that 
\begin{equation}
\label{eqn:BnormCnorm}
CCF_{loc}^{B16} = -CCF_{loc}^{C16} + \beta_{lc}, 
\end{equation}
where $CCF_{loc}^{C16}$ is the local CCF obtained following C16. Since the continuum is a free parameter in the Gaussian fits, the ratio of the areas is still equal to that in Eq.~\ref{eqn:Arat}.

As such, the technique implemented by B16 represents a circular argument. Moreover, if one had the broadband photometry necessary for the correct spectral normalisation, then the planet radii could be determined directly from the light curves alone. 

We stress that a solely brightness-based approach to transmission spectroscopy uses only the equivalent width,
and therefore precludes any retrieval of information on exoplanet radii. It is the inclusion of the spectral dimension that is necessary to determine $R_p$ (i.e. utilising the Doppler information, as is done in \citealt{cameron10, digloria15}). 

\vspace{-15pt}   
\section{Systematic effects on chromatic radius measurements}
\label{sec:sim_recover}
To try to understand how B16 obtained results mimicking Rayleigh scattering, we simulated the transit of HD\,189733\,b and applied their technique. To demonstrate that we could recover our model inputs, we also present results wherein we applied the normalisation from C16 with the correct planet radii for each passband; in doing so we discovered the approximations for $\beta$ in B16 underestimated the planet radii, and thus we also explored a numerical approach for calculating this brightness ratio that was more accurate than the approximation used in B16. 

\vspace{-10pt}
\subsection{Analytical brightness ratio approximation}
\label{subsec:analytical}
Unfortunately, integrating the limb darkened brightness underneath a planet lying off stellar disc centre is not straightforward and even approximate analytical expressions are quite complex (especially if considering ingress and egress regions). For this reason, B16 used the $\beta$ formalism presented in \cite{cameron10}, which was based on the analytical approximations of \cite{ohta05} for the RM effect. Therein the brightness ratio for a fully in-transit planet (i.e. no ingress or egress regions were considered) was defined, under the standard linear limb darkening law, as:

\begin{equation}
\label{eqn:beta}
\beta \approx \Bigg( \frac{R_p}{R_\star} \Bigg)^2 \frac{1-u_1 + u_1*\mu}{1-u_1/3},
\end{equation}
where $R_\star$ is the stellar radius, $R_p$ is the planet radius, $u_1$ is the linear limb darkening coefficient, and $\mu$ is the centre-to-limb planet position. We note that $\mu = \cos(\theta) = \sqrt{1-x_p^2-y_p^2}$, where $\theta$ is the centre-to-limb angle, and $x_p, y_p$ is the centre of the planet \citep[see][for details]{cameron10, cegla16b}. We also note that \cite{ohta05} state that the accuracy of this approximation diminishes with increasing $R_p/R_\star$, arguing that the additional terms in the analytical solution contribute to $\sim$1\% if $R_p/R_\star$ is $\le$ 0.1 and up to few percent if $R_p/R_\star \sim 0.3$. Given that $R_p/R_\star$ is only predicted to vary a few precent in wavelength for particular atmospheric characteristics, such as Rayleigh scattering, this approximation may inject systematic errors that could be misinterpreted as having a physical origin (even if the light curve normalisation is done correctly). 

\vspace{-10pt}
\subsection{Numerical brightness ratio approximation}
\label{subsec:numerical}
To investigate the impact of the accuracy of Eq.~\ref{eqn:beta}, we decided to calculate $\beta$ numerically. For this approach, the in-transit starlight blocked by the planet is still defined as $\beta = F_{loc}/F_\star$, with $F_\star$ and $F_{loc}$ defined as the fluxes of the total stellar disc and the stellar disc under the planet, respectively. Note the observed brightness of the un-occulted star can be analytically determined by integrating a given limb darkening law over the projected stellar disc; for a linear limb darkening this is

\begin{center}
\begin{figure}[t!]
\centering
\includegraphics[scale=0.48, trim = 0.7cm 0cm 0cm 0cm]{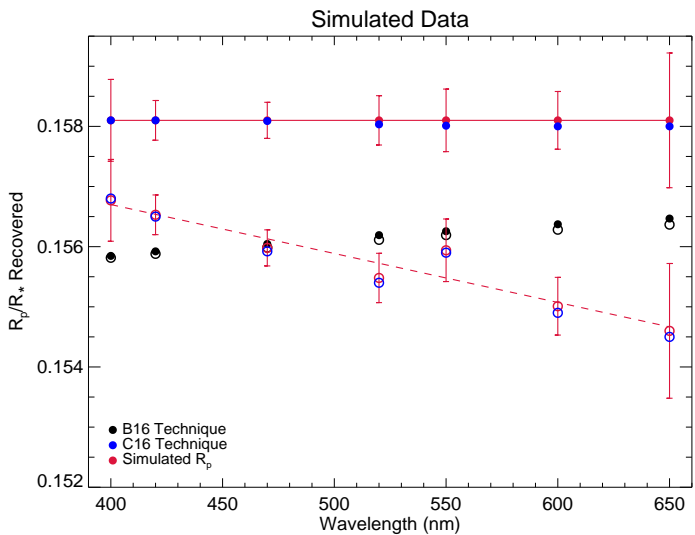}
\caption[]{Recovered average planet radius, $R_p$, from the simulated data for each passband as a function of wavelength (plotted using the middle wavelength in the passband). Filled circles represent when the simulated $R_p$ was constant, and hollow circles represent when the simulated $R_p$ varied; in both cases these are shown in red with the error bars reported by B16 for comparison purposes only. Results from the B16 procedure are shown in black, while those from the numerical approximation herein and the C16 formulation are in blue. The lines represent linear fits to the simulated $R_p$ shown for viewing ease. 
\vspace{-5pt}
} 
\label{fig:Rp_wl}
\end{figure}
\end{center}

\vspace{-35pt}
\begin{equation}
\label{eqn:Idisc_full}
F_\star = R_\star^2 \int_0^{2\pi} \int_0^1 I(\mu) \ \mu \ d\mu \ d\phi = \pi R_\star^2 \ \Bigg(1 - \frac{u_1}{3}\Bigg),
\end{equation}
where $\phi$ is the azimuthal angle. As previously stated, calculating the flux behind the planet analytically is not trivial. Hence, we calculated $F_{loc}$ numerically by constructing a square stellar grid with a width of $2R_p$ centred about the planet position ($x_p, y_p$), with $n$ equal steps in the vertical and horizontal direction. Contributions from steps that did not lie beneath the planet and/or on the stellar disc were excluded. Thus, we approximated the flux behind the planet as
\begin{equation}
\label{eqn:Fp}
F_{loc} \approx \sum I_{xy} \ \Bigg( \frac{2R_p}{n} \Bigg)^2,
\end{equation}
where $I_{xy}$ is the limb darkened intensity at a given position in the aforementioned grid and $(2R_p/n)^2$ is the corresponding area.

Our aim was to try to recover $R_p$ as a function of wavelength, wherein the injected $R_p$ was only used to construct the correct light curves (acting as if we had simultaneous multi-colour photometry). Hence, when trying to recover $R_p$, we started with the broadband planet radius and then allowed it to vary by up to $\pm$ $0.005R_\star$ in steps of 0.0001$R_\star$. The recovered planet radius then corresponded to the planet radii that minimised the difference between $\beta$ and $A_{loc}/A_{out}$. 

\begin{center}
\begin{figure}[t!]
\centering
\includegraphics[scale=0.48, trim = 0.7cm 0cm 0cm 0cm]{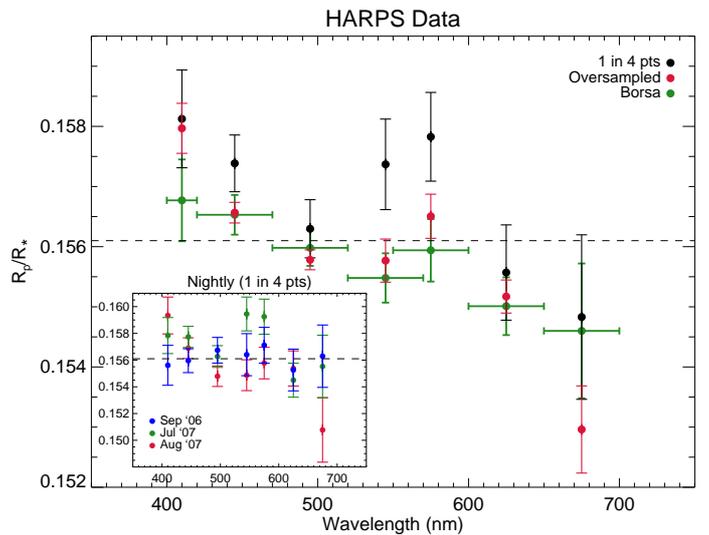}
\caption[]{Recovered average planet radius, $R_p$, for each passband as a function of wavelength for the observed HARPS data. Results from B16 are shown in green (where the wavelength error bars represent the passband wavelength region), and those from our reanalysis are shown in red when using the oversampled CCF and in black when using only every one in four points in the CCF. Subplot: the nightly recovered $R_p$ when using only every one in four points in the CCF (night indicated by colour). The horizontal dashed lines show the mean $R_p$ recovered by the B16 method on the simulated data (i.e. the solid black points in Fig.~\ref{fig:Rp_wl}).
\vspace{-5pt}
} 
\label{fig:Rp_wl_data}
\end{figure}
\end{center}

\vspace{-40pt}
\subsection{Simulated star-planet system}
\label{sec:sim}
We used the simulated stellar grid of \cite{cegla15,cegla16a} and injected into each grid cell a Gaussian profile with a full-width half-maximum (FWHM) of 5~km~s$^{-1}$ (note this width is similar to the expected value for the stellar photosphere). In the simulated star we did not consider any astrophysical effects (i.e. granulation or starspots etc.) other than rigid body stellar rotation, which was set to the value obtained by C16, 3.25~km~s$^{-1}$. We also assumed an edge-on ($i_\star = 90 \degree$) aligned orbit. 

The transit was sampled in 21 equal steps in phase from $-0.02 - 0.02$, centred about mid-transit, with an additional sample at phase = 0.03 to serve as a completely out-of-transit reference. We simulated a transit for each of the seven passbands (from $400 - 700$~nm) used in B16, and applied a linear limb darkening using the coefficients (for each passband) these authors provided. For each of the seven passband transits we injected a planet with a constant radius equal to the value assumed by B16 for the whole HARPS passband ($R_p = 0.1581 R_{\star}$, hereafter referred to as the broadband $R_p$), but varied the limb darkening accordingly.

For this set of transits, we tested the impact of the transit light curve normalisation. In the first case, we followed the procedure in B16, and in the second case we normalised the data following C16 and used the numerical approximation in Sect.~\ref{subsec:numerical} to estimate $R_p$. Examining the first case allowed us to examine any errors introduced using the $\beta$ approximation and/or the B16 normalisation. On the other hand, the second case offered a test case to ensure we could recover the model inputs. 

For a second test, we repeated the above, but varied the radius of the simulated planet; for this we selected $R_p$ equal to the values reported by B16 for each passband. Again we tested two cases: first following the B16 procedure (where the light curve limb darkening varies in each passband, but the light curve radius remains fixed at the broadband $R_p$), and the second using the C16 normalisation (we note the assumed light curve has the correct $R_p$ here) and the numerical approximation in Sect.~\ref{subsec:numerical}.  

\vspace{-10pt}
\subsubsection{Obtaining the planet radius and systematic errors}
\label{sec:recover}
We examined the recovered $R_p$ as a function of stellar disc position, and found only a slight dependance on disc position when following B16. However, if one point each at the ingress and egress regions were included then the dependence on disc position was strong, and including such data would systematically decrease the recovered $R_p$ (as the $\beta$ formulation is not valid in these regions).

Moreover, we found that, regardless of the B16 or C16 normalisation, using the $\beta$ approximation always underestimated the limb darkening behind the planet and therefore also underestimated the true planet radius. This is because the analytical approximation assumes the limb darkening behind the planet is constant, and equal to the value behind the centre of the planet. In reality, the stellar photosphere behind the planet exhibits a range of limb darkening. This is why the numerical model in Sect.~\ref{subsec:numerical} is necessary to recover the $R_p$ injected into the simulated data.

The $R_p$ reported in B16 comes from averaging together the planet radii recovered across the stellar disc. If the limb darkening effects are sufficiently removed (and the stellar profile is constant), then this provides a good means to boost the signal-to-noise in the reported $R_p$. In Fig.~\ref{fig:Rp_wl}, we present the average recovered $R_p$ as a function of wavelength from the simulations, for both tests (when $R_p$ was constant and when it varied). As expected from Sect.~\ref{subsec:tran}, the B16 procedure always results in nearly the same $R_p$, regardless of whether the true $R_p$ varied or not. 

For our numerical approach and the C16 normalisation, we demonstrate accurate recovery of $R_p$ (regardless of whether or not we include ingress and egress data), but only if the light curve normalisation is done with the correct $R_p$ for each passband (using the broadband $R_p$ for all passbands meant only the broadband $R_p$ was recovered). Hence, regardless of the normalisation (i.e. B16 or C16) or the brightness formulation (i.e. $\beta$ or our numerical approximation), we could only retrieve the parameters injected into the system via the transit light curve normalisation, as expected from Sect.~\ref{sec:calc}.

\vspace{-10pt}
\subsection{Reanalysis of the HARPS data}
\label{sec:recover_data}
Our application of the B16 procedure on the simulated data cannot explain the wavelength-dependent planet radii reported in B16. To further investigate this aspect, and to ensure we have applied the B16 method correctly, we have reanalysed the same three transits of HD\,189733\,b following their technique, but using the Levenberg-Marquardt least-squares minimisation from \texttt{MPFIT} \citep[][and references therein]{markwardt09} rather than \texttt{IDL}'s \texttt{GAUSSFIT}\footnote{\texttt{MPFIT} did not produce significantly different results compared to \texttt{IDL}'s \texttt{GAUSSFIT}, but it did allow us to propagate our errors more thoroughly (see C16 for details).}. The results are plotted in Fig.~\ref{fig:Rp_wl_data}, alongside those from B16\footnote{We note that we used the same transit parameters as B16, but there is a slight difference in the template mask used to obtain the CCFs. B16 used the archival data available from the ESO website, where 2 nights used the G2 mask and 1 night used the K5 mask, whereas our data always used the K5 mask. However, this difference is unlikely to impact the analysis since each night had its own master $CCF_{out}$.}. We demonstrate we can reproduce (red points in Fig.~\ref{fig:Rp_wl_data}) results in 1-2$\sigma$ agreement with B16 (in green); hence, we are confident we have applied their technique properly (in both the simulated and observed data). However, we argue that with the correct treatment of the uncertainties the recovered planet radii (in black) are consistent with a flat line (within $1-3\sigma$ of the mean $R_p$ recovered in the simulated data, i.e. the solid black points in Fig.~\ref{fig:Rp_wl}), as expected from Sect.~\ref{sec:calc}. We believe the reason B16 report a trend with wavelength, and we do not, is largely due to differences in our error analysis and Gaussian fitting techniques. 

In B16, the recovered $R_p$ for each stellar disc position and all three transits were averaged together to provide one $R_p$ for each passband, and the reported errors came from the rms of these individual planet radii (i.e. the standard deviation divided by the square root of the total number in the passband). In our analysis, we report the weighted mean for each passband, with the weights being the inverse square of the error for each individual planet radii (where the error was calculated by propagating the errors on the CCF areas as reported from the Gaussian fits following C16, and assuming negligible error on the limb darkening and stellar disc positions). The error on the weighted mean then was simply the square root of the inverse sum of the weights squared. If the errors on individual $R_p$ were all exactly equal to the standard deviation, then the two approaches would yield the same result. 

In addition to this slight difference in error analysis, B16 also applied their Gaussian fits to the oversampled CCF grid provided by the HARPS pipeline (Borsa, Private Comm.). We caution against such an approach, as the oversampling will lead to a significant underestimation of the errors. Hence, we also fit Gaussians to data composed of every one in four points from the original CCFs (to compensate for the original sampling rate of 0.25\,km\,s$^{-1}$ for a mean pixel width of 0.82\,km\,s$^{-1}$); these results are shown in black in Fig.~\ref{fig:Rp_wl_data}.

\begin{table}[t!]
\caption[]{Best fits to observed data}
\centering
\begin{tabular}{c|c|c|c|c|c|c}
    \hline
    \hline
    Data & Function & BIC & $\chi^2_r$ & Function & BIC & $\chi^2_r$ \\
    \hline  
1 in 4 pts & Flat &  13.9 & 2.0 & Linear &  11.7 & 1.6 \\
Oversamp. & Flat &  65.4 & 10.6 & Linear &  29.1 & 5.0 \\
B16 & Flat &  13.1 & 1.9 & Linear &  5.2 & 0.3 \\
    \hline  
  \end{tabular}
\label{tab:bestfit}
\vspace{-15pt}
\end{table}

To test the significance of a trend in $R_p$ with wavelength, we fitted the data with both a flat line and a linear regression, and calculated the reduced chi-squared, $\chi^2_r$, and the Bayesian Information Criterion (BIC); the results are shown in Table~\ref{tab:bestfit}. We note that even if a wavelength-dependent $R_p$ is found, it does not confirm the B16 technique is valid, as we have already shown it is not mathematically possible to retrieve radius variations. Rather, it would serve as a red flag that we do not fully characterise the interplay of the various complexities present in the observations. In particular, stellar activity can alter the observed stellar line shapes and their equivalent widths -- which in turn could lead to spurious radius variations following Sect.~\ref{sec:calc}. Since HD\,189733 is a known active star this is likely scenario; and in agreement with Fig.~5 from B16, wherein the single night analysis with the most apparent slope, July 2007, is also the most magnetically active \citep{cegla16b}. Moreover (and as noted by B16), \cite{mccullough14} have argued that the apparent wavelength dependency in their independent observations of this system are best explained by un-occulted starspots rather than the planet atmosphere.

When using the oversampled CCFs, both our analysis and B16's indicate a slight improvement in fit for the model with a wavelength-dependent slope. However, we find a much worse fit to the data than that found with the B16 results. The high $\chi^2_r$ from our reanalysis indicates an underestimation of the uncertainty in the data, as one would expect when using the oversampled CCFs.  We cannot explain the very low $\chi^2_r$ for the B16 wavelength-dependent fit, which indicates the model is overfitting the data. We note these tests were only performed on our reanalysis of the oversampled data for comparison with B16; for our conclusions on the best-fit, we refer the reader to the analysis on the CCFs sampled every one in four points.

For the properly sampled dataset, we found only a marginal improvement in the fit for the wavelength-dependent model, and do not deem this improvement to be statistically significant (see Table~\ref{tab:bestfit}). Moreover, the best-fit flat model ($R_p$ = 0.1569 $\pm$ 0.0003) lies within 3$\sigma$ of the mean $R_p$ predicted by the simulations. The slight improvement for the sloped model is also heavily influenced by only a couple data points from a single transit, in August 2007, as shown in the subplot of Fig.~\ref{fig:Rp_wl_data}. If the best-fit model is robust, it should withstand removing the August transit; however, doing so means the data is then best-fit by a flat line ($\chi^2_r$ = 1.6, BIC = 11.1 and $\chi^2_r$ = 1.7, BIC = 12.6 for flat and sloped line, respectively).  Consequently, we believe B16 likely report a wavelength-dependent trend in $R_p$ due to insufficient error analysis, and that its agreement with the literature may be purely coincidental.  

\vspace{-10pt}
\section{Conclusions}
\label{sec:conc}
We outline our conclusions on a point-by-point basis below. 

\begin{itemize}

\item The technique presented in B16 using the ratio of the areas of the local (starlight behind the planet) to the out-of-transit CCF cannot be used to determine $R_p$, as $R_p$ must be known a priori for the transit light curve normalisation required for ground-based spectra. This is shown both analytically and using a simulated star-planet system.

\item The analytical $\beta$ approximation used in B16 also introduces (slight) systematic trends with planet position due to inadequately accounting for limb darkening and fractional area occultation effects, and underestimates the value of $R_p$. 

\item We postulate that the $R_p$ variations reported in B16 are likely due to underestimated errors (largely originating from use of oversampled CCFs), as our reanalysis of the HD\,189733\,b transits further demonstrates that the only $R_p$ recoverable is that injected into the transit light curve normalisation. 

\item Chromatic RM measurements from ground-based spectra are not possible without taking the Doppler information into account. Hence, for future measurements, we advise readers to either follow the works of \cite{snellen04, digloria15} or to apply the line-profile tomography of \cite{cameron10} directly on each spectral passband.

\end{itemize}

\begin{acknowledgements} 
We thank the referees, I.~A.~G. Snellen and S. Albrecht, for their careful reading and constructive comments, which improved the clarity of the manuscript. We also thank F. Borsa for useful discussions. Additionally, HMC thanks E. de Mooij for suggesting we use simulated stars to test the work herein. HMC and CAW gratefully acknowledge support from the Leverhulme Trust (grant RPG-249). HMC, VB, CL and AW acknowledge the financial support of the National Centre for Competence in Research ``PlanetS'' supported by the Swiss National Science Foundation (SNSF). CAW also acknowledges support from STFC grant ST/L000709/1, and AW acknowledges additional financial support directly from the SNSF. This research has made use of NASA's Astrophysics Data System Bibliographic Services.
\end{acknowledgements}

\bibliographystyle{aa}
\bibliography{abbrev,mybib}

\end{document}